\documentclass[runningheads]{llncs}
\usepackage[utf8]{inputenc}
\usepackage[T1]{fontenc}
\usepackage{graphicx}
\usepackage{floatrow}
\newfloatcommand{capbtabbox}{table}[][\FBwidth]

\usepackage{multirow}
\usepackage{hyperref}

\usepackage[figurename=Figure]{caption} 

\usepackage{changes}
\usepackage{ulem}
\usepackage{xcolor}
\usepackage{xspace}
\usepackage{soul}
\makeatletter 
\def\bluewave{\bgroup \markoverwith{\lower3.5\p@\hbox{\sixly \textcolor{blue}{\char58}}}\ULon}
\font\sixly=lasy6 
\makeatother 

\usepackage{algorithm}
\usepackage{stmaryrd}
\usepackage{amsmath}
\usepackage{amssymb}

\usepackage[noend]{algpseudocode}

\algblock{Input}{EndInput}
\algnotext{EndInput}
\algblock{Output}{EndOutput}
\algnotext{EndOutput}
\newcommand{\Desc}[2]{\State \makebox[2em][l]{#1}#2}
\algnewcommand{\LeftComment}[1]{\Statex \(\triangleright\) #1}

\usepackage{subfigure}
\usepackage{graphicx}
\usepackage{makecell}

\newcommand{\tool}{Limited Ltd.\xspace}
\newcommand{\icmprl}{ICMP rate limiting\xspace} 

\usepackage{balance}
\author{Kevin Vermeulen\inst{1}
\and
Burim Ljuma\inst{1}
\and
Vamsi Addanki\inst{1}
\and
Matthieu Gouel\inst{1} \and\\
Olivier Fourmaux\inst{1}
\and
Timur Friedman\inst{1}
\and
Reza Rejaie\inst{2}
}
\authorrunning{Vermeulen et al.}
\institute{Sorbonne Université \and University of Oregon}

\begin{document}
\title{Alias Resolution Based on ICMP Rate Limiting}
%
%
%
\maketitle              
%
\begin{abstract}
Alias resolution techniques (e.g., \textsc{Midar}) associate, mostly through active measurement, a set of IP addresses as belonging to a common router.  These techniques rely on distinct router features that can serve as a signature. Their applicability is affected by router support of the features and the robustness of the signature. This paper presents a new alias resolution tool called \tool that exploits \icmprl, a feature that is increasingly supported by modern routers that has not previously been used for alias resolution. It sends ICMP probes toward target interfaces in order to trigger rate limiting, extracting features from the probe reply loss traces. It uses a machine learning classifier to designate pairs of interfaces as aliases. We describe the details of the algorithm used by \tool and illustrate its feasibility and accuracy. \tool not only is the first tool that can perform alias resolution on IPv6 routers that do not generate monotonically increasing fragmentation IDs (e.g., Juniper routers) but it also complements the state-of-the-art techniques for IPv4 alias resolution.
All of our code and the collected dataset are publicly available.
\end{abstract}

\section{Introduction}\label{sec:intro}
Route traces obtained using {\small \texttt{traceroute}} and similar
tools provide the basis for generating maps that reveal the inner structure of the Internet's many
autonomously administered networks, but not necessarily at the right
level of granularity for certain important tasks. Designing
network protocols~\cite{willinger2009mathematics} and understanding
fundamental properties of the Internet's
topology~\cite{gunes2007importance} are best done with router-level
maps.  Rather than revealing routers, {\small \texttt{traceroute}}
only provides the IP addresses of individual router interfaces. The
process of grouping IP addresses into sets that each belong to a
common router is called \textit{alias resolution}, and this paper
advances the state of the art in alias resolution.

A common approach to alias resolution is to send probe packets to IP
addresses, eliciting reply packets that display a feature that is
distinctive enough to constitute a signature, allowing replies coming
from a common router to be matched. This paper describes a new type of
signature based upon a functionality, \textit{\icmprl}, in
which an Internet-connected node (router or end-host) limits the
ICMP traffic that it sends or receives within a certain window of
time. This new signature enjoys much broader applicability
than existing ones for IPv6 alias resolution, thanks to \icmprl being
a required function for IPv6 nodes. The signature also complements IPv4 existing signatures.

Our contributions are: (1) The \tool\ algorithm, a new
signature-based alias resolution technique that improves 
alias resolution coverage by 68.4\% on Internet2 for IPv6 and by 40.9\% on \textsc{Switch} for IPv4
(2) a free, open source, and
permissively licensed tool that implements the algorithm.

We evaluate \tool\ by comparing its performance to two
state-of-the-art alias resolution tools:
Speedtrap~\cite{luckie2013speedtrap} for IPv6, and
\textsc{Midar}~\cite{keys2013internet} for IPv4,  
using ground
truth provided by the Internet2 and \textsc{Switch} networks.

The remainder of this paper is organized as follows:
Sec.~\ref{sec:related-work} provides technical background and related
work for both alias resolution and ICMP rate
limiting. Sec.~\ref{sec:algorithm} describes the \tool\
technique in detail. Sec.~\ref{sec:evaluation} presents the
evaluation. Sec.~\ref{sec:ethical} discusses ethical considerations and  Sec.~\ref{sec:conclusion} summarizes our conclusions and
points to future work.
	
\section{Background and Related Work}\label{sec:related-work}

\tool\ is the latest in a long line of alias resolution methods
stretching back over twenty-plus years.
An inventory of all previously known techniques (Table~\ref{table:alias-techniques})
shows that there are only four techniques known to work for IPv6. Of 
these, there is a publicly-available tool for only one: Speedtrap~\cite{luckie2013speedtrap}. But Speedtrap has a known limitation of only working 
on routers that generate monotonically increasing IPv6 fragmentation IDs, whereas there is an entire class of routers, such as those from Juniper, that do not generate IDs this way. Relying upon monotonically increasing IP IDs for IPv4, as does state-of-the-art \textsc{Midar}~\cite{keys2013internet}, presents a different issue: fewer and fewer routers treat IPv4 IP IDs this way due to a potential vulnerability~\cite{ensafi2014detecting,private-caida}.
\tool is a publicly available tool that does not rely upon monotonically increasing IDs, thereby enabling IPv6 alias resolution on Juniper routers for the first time and IPv4 alias resolution on a growing class of routers for which \textsc{Midar} will no longer work.
\begin{table*}[t]
\tiny
\centering
\resizebox{\linewidth}{!}{%
\begin{tabular}{|c|c|c|c|c|c|}
\hline
\multirow{2}{*}{\makecell{Year}} & \makecell{Basis} & \multirow{2}{*}{\makecell{Algorithms \\ and tools}} & \multirow{2}{*}{\makecell{Condition of applicability}} & \makecell{IPv4} & \makecell{IPv6}  \\
\cline{5-6} & \makecell{(\textbf{s}) = signature \\ (\textbf{t}) = topology \\ (\textbf{o}) = other} & & & \multicolumn{2}{l|}{\makecell{($\boldsymbol{\tau}$) = tool \\ ($\boldsymbol{\delta}$) = dataset}} 	 \\
\hline
\multirow{2}{*}{\makecell{1998~\cite{pansiot1998routes}}} & \multirow{2}{*}{\makecell{source \\ IP address~(\textbf{s})}} & \makecell{Pansiot \\ and Grad~\cite{pansiot1998routes}}   					 & \makecell{respond with a common \\ IP address in} & \makecell{yes}                                                                                                                      &                                                                                                                                                                                                          \\  
\cline{3-3}\cline{5-6}
                                                          												 &											& \makecell{\textit{Mercator}~\cite{govindan2000heuristics}}                 &  \makecell{ICMP Destination \\ Unreachable messages}																																									  & \makecell{yes \\ ($\boldsymbol{\tau}$)~($\boldsymbol{\delta}$)}                                                                            &                                                						           						    \\	        
\hline
        \multirow{3}{*}{\makecell{2002~\cite{Spring:2002:MIT:964725.633039}}} & \multirow{3}{*}{\makecell{IP ID~(\textbf{s})}} & \makecell{\textit{Ally}~\cite{Spring:2002:MIT:964725.633039}} & \multirow{3}{*}{\makecell{send replies with a shared \\ IP ID counter \\ that increases monotonically \\ with each reply}} & \makecell{yes \\ ($\boldsymbol{\tau}$)}                                                                                               &                                                        \\
\cline{3-3}\cline{5-6}
                                                                                                                                      &                                                                              & \makecell{\textit{RadarGun}~\cite{bender2008fixing}}                &                                                                                                                                                                                                   & \makecell{yes \\ ($\boldsymbol{\tau}$)}                                                                                               &  										  			                                                             \\
\cline{3-3}\cline{5-6}                                                                                                                
                                                                                                                                      &                                                                              & \makecell{\textit{\textsc{Midar}}~\cite{keys2013internet}}          &																																											    & \makecell{yes \\ ($\boldsymbol{\tau}$)~($\boldsymbol{\delta}$)}																	&								                                              \\
\hline
	\makecell{2002~\cite{Spring:2002:MIT:964725.633039}} & \makecell{Reverse DNS~(\textbf{o})} & \makecell{\textit{Rocketfuel}~\cite{Spring:2002:MIT:964725.633039} \\ \textit{\textsc{Aroma}}~\cite{kim2007efficient}} & \makecell{IP address resolves to a name} & \makecell{yes} & \makecell{yes}   \\
\hline
     \multirow{2}{*}{\makecell{2006~\cite{gunes2009resolving}}} & \multirow{2}{*}{\makecell{\texttt{traceroute}~(\textbf{t})}} & \makecell{\textit{\textsc{Apar}}~\cite{gunes2009resolving}}  & \multirow{2}{*}{\makecell{respond with ICMP Time \\ Exceeded messages}} & \makecell{yes}         								&    \\
\cline{3-3}\cline{5-6}     
																								 &																							 & \makecell{\textit{kapar}~\cite{keys2010internet}}			       &																						 				 		  & \makecell{yes \\ ($\boldsymbol{\tau}$)~($\boldsymbol{\delta}$)} &  \\
\hline
	\makecell{2010~\cite{sherry2010resolving}} & \makecell{IP Prespecified \\Timestamp \\ option~(\textbf{s})} & \makecell{Sherry et al.~\cite{sherry2010resolving} \\ \textit{Pythia}~\cite{marchetta2013pythia}} & \makecell{fill in timestamps \\ as specified \\ by the option} & \makecell{yes} &  \\
\hline
	\makecell{2010~\cite{qian2010route}} & \makecell{IPv6 source \\ routing~(\textbf{s})} & \makecell{Qian et al.~\cite{qian2010route,qian2010utilizing}} & \makecell{source routing \\ must be enabled} & & \makecell{yes}   \\
\hline
	\makecell{2013~\cite{luckie2013speedtrap}} & \makecell{IPv6 fragmentation \\ identifier~(\textbf{s})}  & \makecell{\textit{{Speedtrap}}~\cite{luckie2013speedtrap}} & \makecell{IDs elicited from responses \\ increase monotonically} & & \makecell{yes \\ ($\boldsymbol{\tau}$)~($\boldsymbol{\delta}$)} \\
\hline
	\makecell{2013~\cite{sherwood2008discarte}} & \makecell{IP Record Route \\ option~(\textbf{t})} & \makecell{\textit{{DisCarte}}~\cite{sherwood2008discarte}} & \makecell{fill in IP addresses \\ as specified by the option} & \makecell{yes} & \\
\hline
	\makecell{2015}~\cite{padmanabhan2015uav6} & \makecell{IPv6 unused \\ address~(\textbf{s})} & \makecell{Padman-\\abhan et al.~\cite{padmanabhan2015uav6}} & \makecell{126 prefixes on a point \\ to point link} & & \makecell{yes}   \\
\hline
	\makecell{2019} & \makecell{ICMP rate \\ limiting~(\textbf{s})} & \makecell{\textit{\tool}} & \makecell{\icmprl shared by \\ interfaces of the router} & \makecell{yes \\ ($\boldsymbol{\tau}$)~($\boldsymbol{\delta}$)} & \makecell{yes \\ ($\boldsymbol{\tau}$)~($\boldsymbol{\delta}$)}   \\
\hline
\end{tabular}}
\caption{Alias resolution methods}
\label{table:alias-techniques}
\end{table*}

Regarding ICMP, the Internet Control Message Protocol: its
IPv4 and IPv6 variants~\cite{postel1981rfc,conta2006internet}
allow routers or end-hosts to send error and informational messages. 
The RFC for ICMPv6~\cite{conta2006internet} cites the ``bandwidth and
forwarding costs'' of originating ICMP messages to motivate the need
to limit the rate at which a node originates ICMP messages. 
It also recommends the use of a token bucket mechanism for rate limiting. 
It explicitly calls for compatibility with \texttt{traceroute} by stating that ``Rate-limiting mechanisms that cannot cope with bursty traffic (e.g., traceroute) are not recommended''. Furthermore, it states that, in the case of ``ICMP messages
[being] used to attempt denial-of-service attacks by sending back to back erroneous IP packets'', an implementation that correctly deploys the recommended token bucket mechanism ``would be protected by the ICMP error rate limiting mechanism''. The RFC makes \icmprl mandatory for all IPv6 nodes. \icmprl is a supported feature on all modern routers but its implementation may vary by vendor~\cite{cisco-rate-limiting,cisco-rate-limiting-icmpv6,cisco-rate-limiting-copp,cisco-rate-limiting-car,juniper-rate-limiting-1,juniper-rate-limiting-2,juniper-rate-limiting-icmpv6,juniper-rate-limiting-policer} based on  ICMP message type and IP version.
\icmprl can be performed on incoming traffic or generated replies. \tool makes no distinction between the two. It works whenever
multiple interfaces of a router are subject to a common \icmprl mechanism, i.e., when there is a shared token bucket across multiple interfaces.
Vendor documentation~\cite{juniper-rate-limiting-1,juniper-rate-limiting-2,juniper-rate-limiting-icmpv6,cisco-rate-limiting-copp}, 
indicates that \texttt{ping} packets are more likely to trigger shared \icmprl
behavior. We validated this observation in a prior survey and in a lab environment.
In particular on Juniper (model J4350, JunOS 8.0R2.8), we observed a 
shared \icmprl mechanism for Echo Reply, Destination Unreachable and Time Exceeded packets across all of its interfaces by default. 
But on Cisco (model 3825, IOS 12.3), we observed that the rates for Time Exceeded and Destination Unreachable packets are limited on individual interfaces by default, and only the rate for Echo Reply packets is shared across different interfaces~\cite{cisco-acl}. 
Therefore, we adopted the \texttt{ping} Echo Request and Echo Reply mechanism in our tool to maximize the chances of encountering shared ICMP rate limits across router interfaces.

A few prior studies have examined \icmprl behavior 
in the Internet. Ravaioli et
al.~\cite{ravaioli2015characterizing} identified two types of behavior
when triggering \icmprl of Time Exceeded messages by an interface: on/off and non on/off. Alvarez et al.~\cite{alvarez2017ietf} demonstrated that ICMP Time Exceeded rate limiting is more widespread in IPv6 than in IPv4.  Guo and
Heidemann~\cite{guo2018detecting} later proposed an algorithm, 
\textsc{Fader}, to detect ICMP Echo Request/Reply rate limiting at
very low probing rates, up to 1 packet per second. They found rate
limiting at those rates for very few /24 prefixes.
%
Our work is the first one that exploits the shared nature of \icmprl across different interfaces of a router as a signature to relate these interfaces for alias resolution. 


\section{Algorithm}
\label{sec:algorithm}
The main intuition behind our approach is that two interfaces of a router that implements shared \icmprl, should exhibit a similar loss pattern if they are both probed by ICMP packets at a cumulative rate that triggers rate limiting. The key challenges are to efficiently trigger rate limiting and reliably associate aliases based on the similarity of their loss patterns despite the noise due to independent losses of probes and replies.

Pseudo code~\ref{algo:overview} describes how {\tool} divides a set of input IP addresses into subsets that should each be an alias set.
It proceeds iteratively, taking the following steps in each iteration:
First, a random IP address from the input set is selected as a {\em seed}, with all remaining members of the input set being {\em candidate} aliases for the seed.
The seed is probed at incrementally higher rates until the rate $r_s$ that induces \icmprl is identified ($\mathtt{find\_rate}()$). Then, the seed is probed at that rate of $r_s$ while all of the candidates interfaces are simultaneously probed at low rates. All probing takes place from a single vantage point. Loss traces for reply packets from the seed and each of the candidate interfaces are gathered. It is very challenging to infer that two interfaces are aliases by directly correlating their loss traces. Instead, the algorithm extracts a set of features from each loss trace and collectively uses these as the signatures of the corresponding interfaces($\mathtt{signatures}()$). 
Using a classification technique ($\mathtt{classify}()$), the algorithm examines whether the signatures of candidate and seed are sufficiently similar to classify them as aliases,
in which case the candidate is added to an alias set ($A_s$). Each identified alias set is refined through further testing in order to reduce the chance of false positives ($\mathtt{refine}()$). Finally, the alias set is removed from the input set, and iterations continue until the input set is empty.
The remainder of this section further details these steps.
\begin{algorithm}[t]
\caption{\texttt{\tool}}\label{algo:overview}
\begin{algorithmic}
  \Input
  \Desc{$S$: a set of IP addresses}
  \EndInput
  \Output
  \Desc{$A$: a set of alias sets}
  \EndOutput
  
  \State $K \leftarrow \mathtt{controls}(S)$ : set of controls
  \State $A \leftarrow \emptyset$
  \While{$S \neq \emptyset$} 
  \State choose at random a seed $s \in S$
   \State $C_s \leftarrow S \setminus \{ s\}:$ candidate aliases for $s$
   \State $r_s \leftarrow \mathtt{find\_rate}(s):$ rate limiting rate for $s$
   \State $\Sigma_s \leftarrow \mathtt{signatures}(s,r_s,C_s, K):$ set of signatures
   \State $A_s \leftarrow \{ s\}:$ alias set for $s$
   \For {$c \in C_s$}: for each candidate $c$
      \State $\sigma_{s,c} \in \Sigma_s$ is the pairwise signature for $s$ and $c$ 
      \If{ $\mathtt{classify}(\sigma_{s,c}) ==
        \mathtt{true}$} 
        \State $s$ and $c$ are aliases
        \State $A_s \leftarrow A_s \cup \{c\}:$ add $c$ to the alias set
        \EndIf
      \EndFor
    \State $A_s \leftarrow \mathtt{refine}(A_s):$ try to reduce false positives
    \State $A \leftarrow A \cup \{A_s\}:$ add the new alias set to $A$
    \State $S \leftarrow S \setminus A_s:$ remove the aliases of $s$ from $S$
  \EndWhile
  \Return $A$  
\end{algorithmic}
\end{algorithm}

\subsection{Triggering \icmprl}
\label{sec:triggering}
The goal of $\mathtt{find\_rate}(s)$ is to efficiently determine $r_s$, the probing rate that triggers \icmprl at the router to which seed $s$ belongs. It proceeds by probing the seed with ICMP Echo Request probes across multiple rounds, increasing the probing rate with each round until the loss rate of observed ICMP Echo Replies enters a target range. The target loss range should be sufficiently large to minimize the effect of random independent losses and also relatively small to minimize the load on the router. To satisfy these two opposing conditions, we empirically set the range at 5 to 10\%.
The probing rate remains constant during each round. The rate is low (64 pps) for the first round, and exponentially increases in consecutive rounds until the loss rate falls within (or exceeds) the target range.\footnote{We have explicitly verified that the actual probing rate is not limited by the network card or other factors.}
If the observed loss rate is within the target range, the probing is concluded and the last rate is reported as $r_s$. But if the loss rate is higher than the target range, up to eight additional rounds are launched in a binary search between the last two rates.
\begin{figure}[t]
\centering
\subfigure{\includegraphics[width=.40\textwidth]{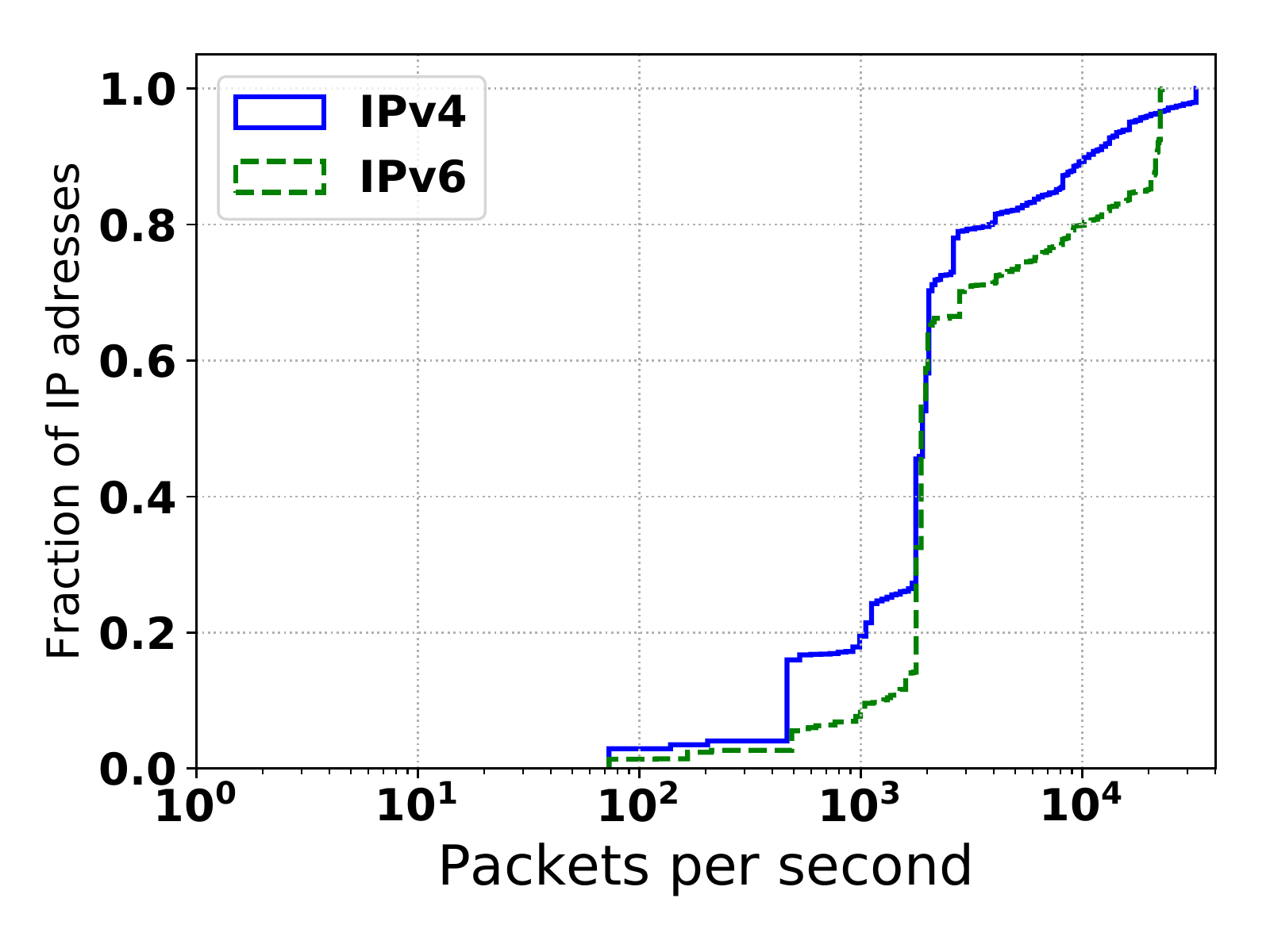}}%
\subfigure{\includegraphics[width=.40\textwidth]{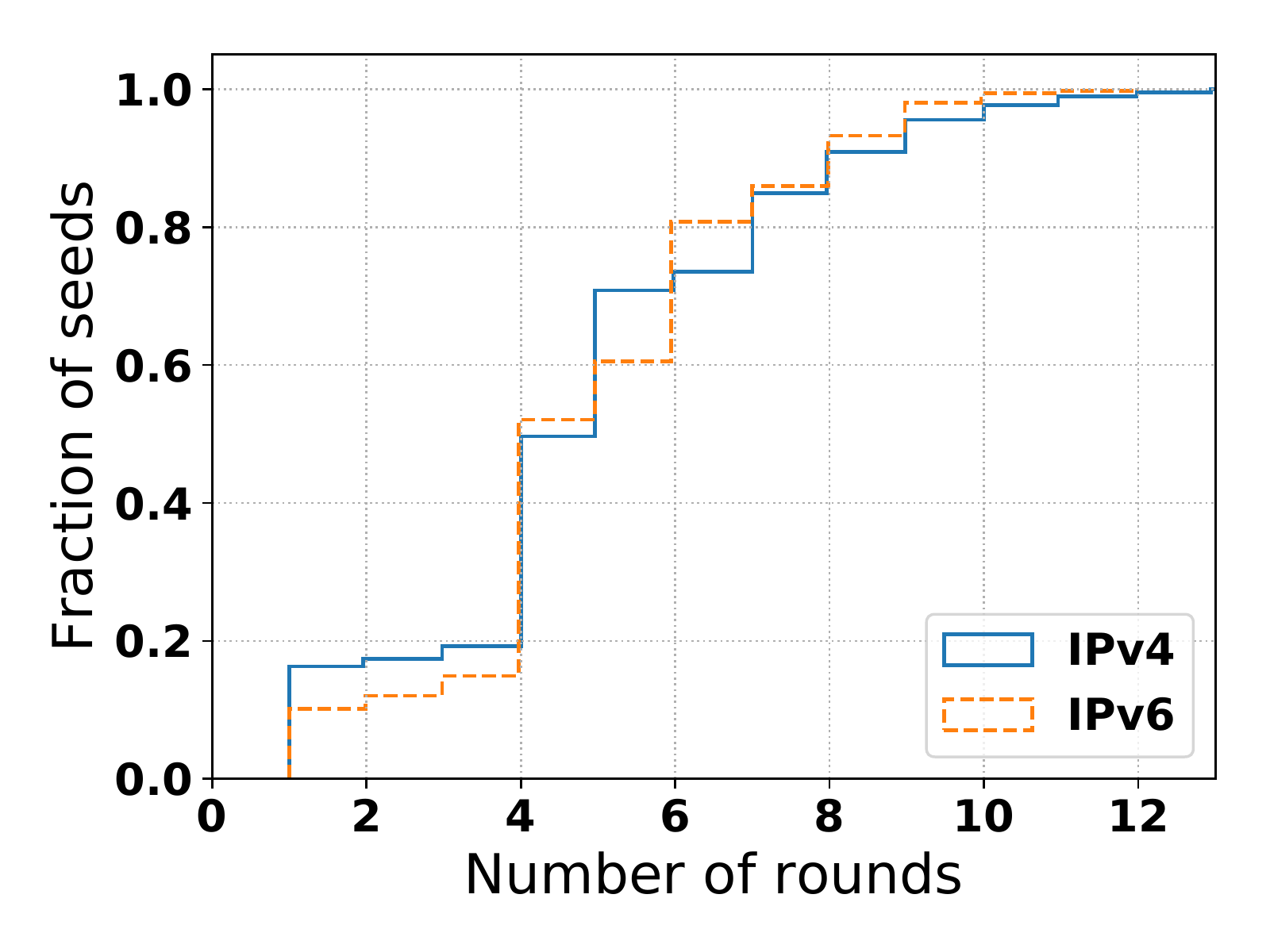}}%
\caption{CDF of the probing rate $r_s$ (left) and the number of probing rounds (right) to trigger \icmprl for 2,277 IPv4 and 1,099 IPv6 addresses.}
\label{fig:algorithm-infos}
\end{figure}
If the loss rate still does not fall within the target range, the probing rate that generates the loss rate closest to the range is chosen.
If the target loss range is not reached as the probing reaches a maximum rate (32,768 pps), the probing process ends without any conclusion.
The duration of each round of probing should be sufficiently long to reliably capture the loss rate while it should also be limited to control the overhead of probing. We experimentally set the duration of each round of probing to 5 seconds, followed by a period, of equal length, of no probing.
%
The right plot of Fig.~\ref{fig:algorithm-infos} presents the CDF of the number of probing rounds to trigger the target loss rate for thousands of IPv4 and IPv6 interfaces (using our dataset from Sec.~\ref{sec:classifier}). We observe that for 90\% of IPv4 or IPv6 interfaces, the \icmprl is triggered in less than 8 rounds of probing. 
The left plot of Fig.~\ref{fig:algorithm-infos} shows the CDF of the probing rate that triggered the target loss rate (i.e., the inferred rate for triggering the \icmprl) across the same IPv4 and IPv6 interfaces. This figure indicates that for 70\% (80\%) of IPv6 (IPv4) interfaces, \icmprl is triggered at less than 2k pps. This result confirms that our selected min and max probing rate covers a proper probing range for more than 99\% of interfaces.
We note that the binary search process failed to reach the target loss rate for fewer than 1\% of the interfaces. All the parameters of our probing strategy are empirically determined. 
Section~\ref{sec:ethical} elaborates on
the ethical considerations associated with 
the probing scheme.

\subsection{Generating interface signatures}
\label{sec:signature}

A signature based on the loss traces of individual interfaces is obtained by probing the seed interface at its target rate ($r_s$) while simultaneously probing each candidate interface at the low rate of $R_c$ pps.
Probing a large number of candidate interfaces in each round may lead to a better efficiency, 
but the aggregate probing rate
should remain low so that it does not independently trigger \icmprl even if all those candidates are in fact aliases.
To address these two constraints, we set the number of candidate interfaces that are considered in each round to 50 and 
$R_c$ to 10~pps. In an unlikely scenario that all of these 50 candidate interfaces are aliases, this strategy leads to a 500~pps probing rate for the corresponding router that does not trigger \icmprl in 90\% of routers, as we showed in the left plot of Fig.~\ref{fig:algorithm-infos}.\footnote{The largest reported alias set by \textsc{Midar} and Speedtrap has 43 interfaces. Therefore, the likelihood of observing 50 candidate interfaces that are all aliases is low.}
\subsubsection{Control Interface.} In order to distinguish the observed losses in the loss traces for the target interfaces (i.e., seed $s$ and individual candidate $c$) that are not related to \icmprl, we also consider another interface along the route to each target interface and concurrently probe them at a low rate (10 pps). These interfaces are called the \textit{controls}, $\kappa_s$ and $\kappa_c$. 
The control $\kappa_i$ for target interface $i$ is identified by conducting a Paris Traceroute~\cite{augustin2006avoiding} towards $i$ and selecting the last responsive IP address prior to $i$.\footnote{\tool maintains the flow ID necessary to reach
$\kappa_s$ in subsequent probing of $s$ and $\kappa_s$.}
The loss rate for $\kappa_i$ also forms part of $i$'s signature. In practice, the \textit{controls} are identified at the beginning of the \tool procedure by conducting route traces to all IP addresses in the input set
$S$. This corresponds to $\mathtt{controls}()$ and $K$ is the resulting set of controls. 

\subsubsection{Inferring Alias Pairs.}
\begin{figure}[t]
\begin{floatrow}
\capbtabbox{%
\resizebox{\linewidth}{!}{%
\begin{tabular}{|l|c|c|c|c|}
\cline{2-5}
    \multicolumn{1}{l|}{} & \makecell{Seed \\$s$} & \makecell{Candidate \\$c$} &
    \makecell{Control \\$\kappa_{s}$}& \makecell{Control \\$\kappa_{c}$} \\
\hline
     \makecell{Loss rate}& x & x & x & x\\
\hline 
     \makecell{Change point}& x & x & &\\
\hline
	\makecell{$\mathtt{gap} \rightarrow \mathtt{gap}$ \\ transition probability}& x & x & & \\
\hline  	
  	\makecell{$\mathtt{burst} \rightarrow \mathtt{burst}$ \\ transition probability}& x & x & & \\
\hline
	\makecell{Pearson correlation \\ coefficient} & \multicolumn{2}{c|}{x} & & \\
\hline
\end{tabular}}}
{%
  \caption{Selected features for a Signature.}%
  \label{table:features}
}
\ffigbox{%
\includegraphics[width=.50\textwidth]{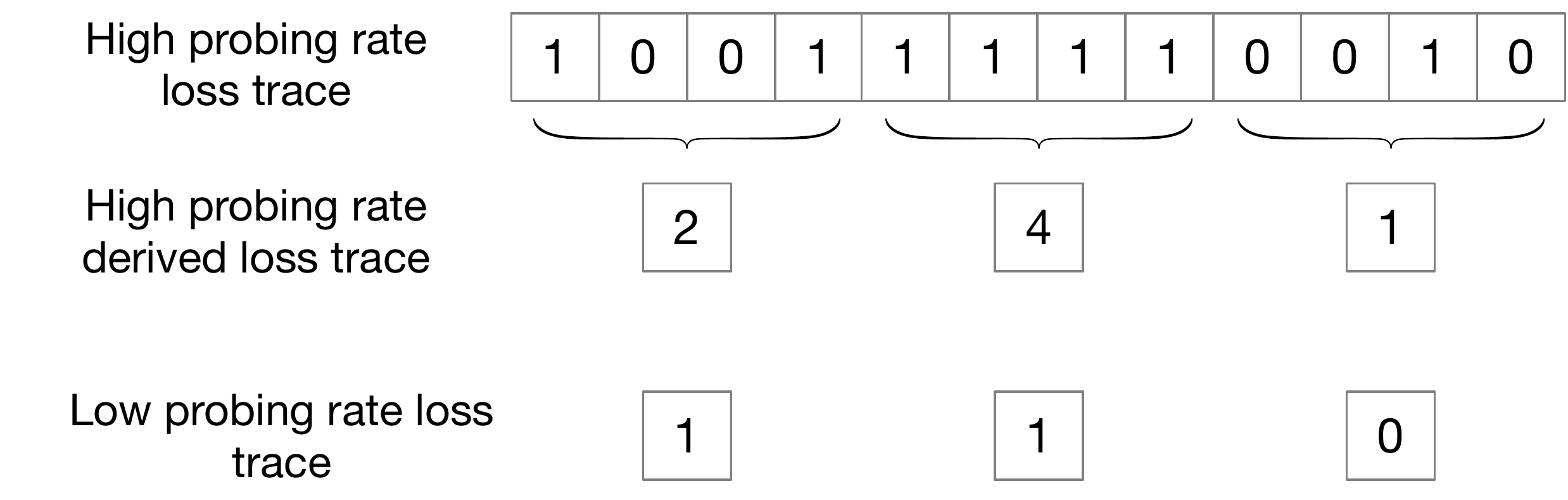}%
}
{
\caption{Mapping between loss traces with different length.}
\label{fig:high-rate-low-rate-correlation}
}
\end{floatrow}
\end{figure}

\begin{figure}[t]
\RawFloats
\begin{floatrow}

\ffigbox{%
\includegraphics[width=.40\textwidth]{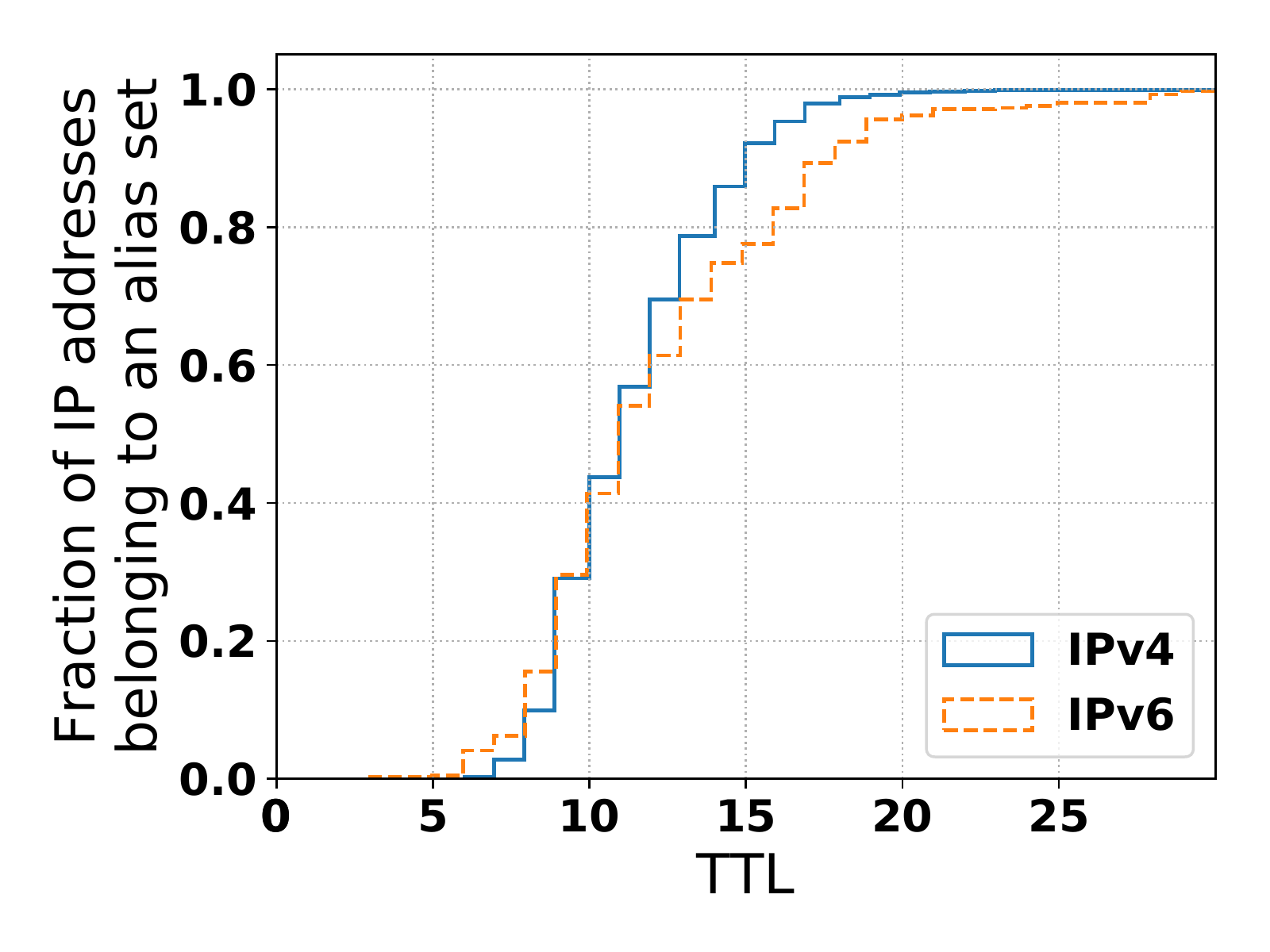}%
}
{
\caption{CDF of the TTL distance from the \tool vantage point of the IP addresses belonging 
to an alias set in our training data.}
\label{fig:ttl-distribution}
}
\ffigbox{%
\includegraphics[width=.40\textwidth]{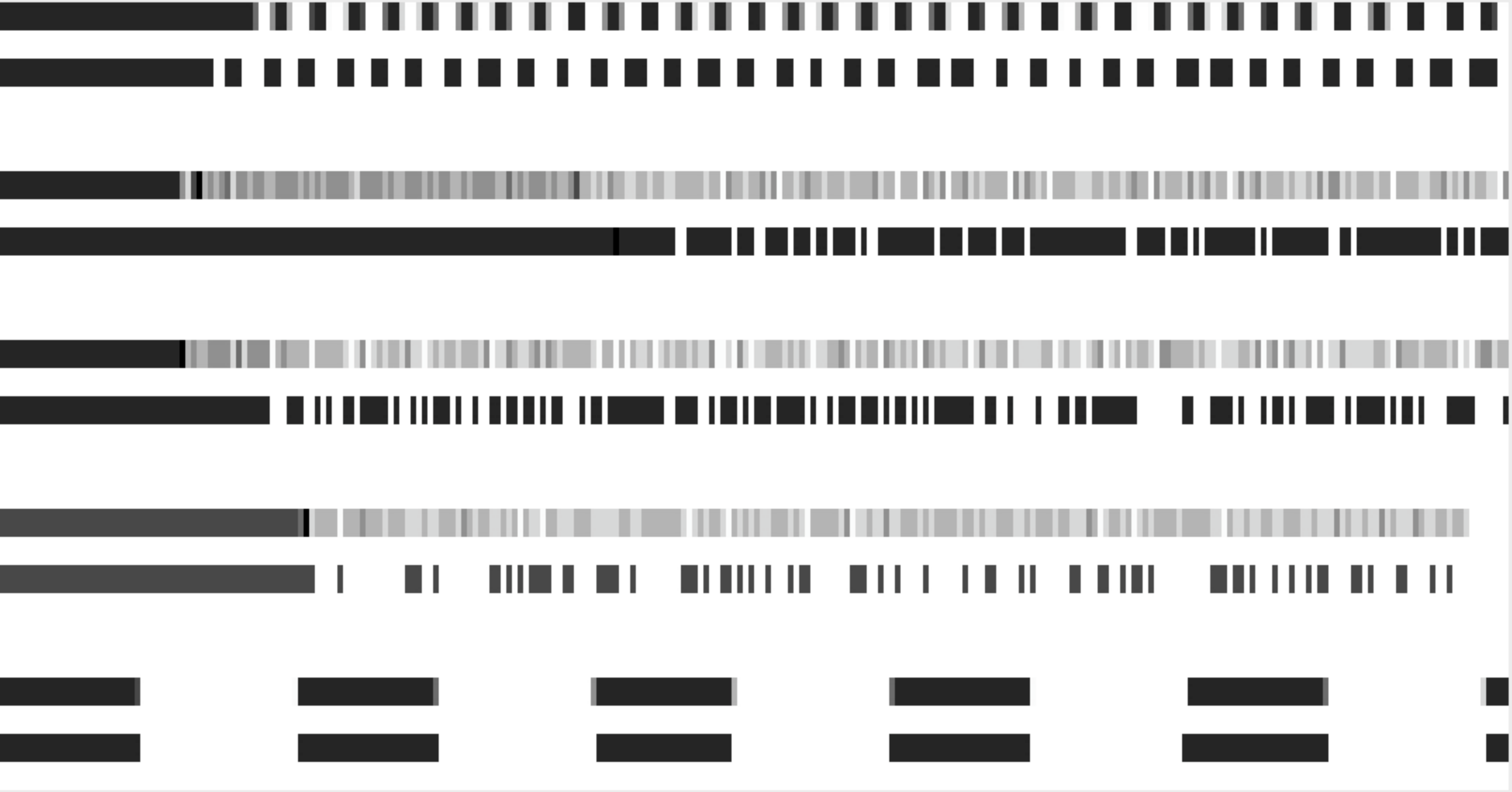}%
}
{
\caption{
Raw times series of loss traces of pairs of aliases.}
\label{fig:raw-traces}
}
\end{floatrow}
\end{figure}

The above probing strategy produces a separate loss trace for each interface. We have found that when losses occur simultaneously at pairs of alias interfaces, they can do so in multiple ways, as the five examples in Fig.~\ref{fig:raw-traces} illustrate.
The black and white strokes in each trace correspond respectively to received and lost ICMP Echo Replies, and their
varied patterns defy attempts to find simple correlations.
We therefore use a machine learning classifier to identify pairs of aliases. It is based on the following features extracted from loss traces that, intuitively, we believe capture the temporal pattern of the losses in each trace. (See also Table~\ref{table:features}.)

\begin{enumerate}
\item{Loss rate:}
This is simply the number of losses in the trace divided by the total
number of probes in the trace.

\item{Change point detection:}
This is the point in a time series (such as our loss traces) when the probability distribution of a time series changes~\cite{aminikhanghahi2017survey}. We adopt
a method based on the variation of the mean and the variance~\cite{changepoint-library}.
\item{Transition probabilities:}
These are obtained by using each loss trace to train a Gilbert-Elliot two-state Markov
model, in which losses occur in the \texttt{burst} state and no losses occur in the \texttt{gap} state. The P($\mathtt{gap} \rightarrow \mathtt{gap}$) and
P($\mathtt{burst} \rightarrow \mathtt{burst}$) transition probabilities are sufficient to fully describe the model since other two probabilities can be easily calculated from these. For example, P($\mathtt{gap}
\rightarrow \mathtt{burst}$) = 1 - P($\mathtt{gap} \rightarrow \mathtt{gap}$).

\item{Correlation coefficient:} The Pearson correlation coefficient between the two loss traces is used as a measure of similarity between them. 
Calculating this coefficient requires both time series to have the same number of values but our loss traces do not meet this condition since we use a higher probing rate for the seed. To address this issue, we condition the seed's loss trace to align it with the loss trace of other interfaces as shown in Fig.~\ref{fig:high-rate-low-rate-correlation}. 
In this example, the length of the loss trace of the seed is four times longer than the ones from the other interfaces. 
We consider groups of four consecutive bits in the seed loss trace and convert it to the sum of the 1's. The resulting loss 
trace has a lower rate and can be directly correlated with other loss traces.
\end{enumerate}

\subsection{Classifying the signatures}
\label{sec:classifier}
We use the \textit{random forest} 
classifier from the \texttt{scikit-learn} Python machine learning library~\cite{scikit-learn}. If it identifies two interfaces as aliases based on their signatures, $\mathtt{classify}()$ returns \texttt{true}; otherwise, \texttt{false}.
%
%
There are several challenges to building such a classifier: (1) it must learn from training data that represents the diversity of possible loss traces generated by pairs of aliases; (2) it should be able to distinguish between losses triggered by \icmprl and unrelated
losses; (3) it should have a high precision, so that \tool minimizes
false positives; and (4) if the training data come from
other alias resolution techniques, such as \textsc{Midar} and Speedtrap, it must be able to generalize to pairs that they
cannot find. We tackled these challenges as follows.
%


\subsubsection{Training and testing data.}
\label{training-data}
We have access to ground truth router-level topology for two networks, Internet2 and \textsc{Switch}, but these do not suffice to capture the richness of router behaviors in the Internet as a whole.
We therefore randomly selected routable IPv4 and IPv6 prefixes from the RIPE registry~\cite{ripe-registry}, and conducted multipath Paris
Traceroute~\cite{vermeulen2018multilevel} from PlanetLab
Europe~\cite{planet-lab-europe} nodes towards the first address in each prefix. This procedure yielded 25,172 IPv4 addresses in 1,671 autonomous systems (ASes) and 18,346 IPv6 addresses in 1,759 ASes from 6,246 and 4,185 route traces, respectively. We use \textsc{Midar} and Speedtrap to identify IPv4 and IPv6 alias sets, respectively, since both tools are known to have low false positive rates. Pairs of interfaces from these sets are used as labeled as \texttt{true}.
For the \texttt{false} labels, we take the conservative approach of selecting pairs of IP addresses that are more than 6 hops from each other in a given route trace. 
The 6 hop value is empirically set, as 99.9\% of the alias pairs identified by \textsc{Midar} and Speedtrap are fewer than 6 hops apart.
This labeling process identified 70,992 unique IPv4 and 7,000 unique IPv6 addresses. 15,747 of IPv4 and 1,616 IPv6 addresses are labeled as aliases forming 
2,277 IPv4 and 1,099 IPv6 alias sets, respectively.
Fig.~\ref{fig:ttl-distribution} shows the CDF of hop count distance between our vantage point and selected IP addresses and indicates that these targets are 7-17 hops away from the vantage point.
For each alias set, one address is chosen at random to play the role of the seed $s$, and the candidate set is composed of all of the other aliases in the set that are rounded up with some randomly selected non-aliases to make a $C_s$ of size between 2 (minimum one alias and one non-alias) and 50 (our cap for the number of addresses to be simultaneously probed at a low rate). The high rate $r_s$ at which to probe the seed is found through $\mathtt{find\_rate}(s)$, and the signatures are generated through $\mathtt{signatures}(s,r_s,C_s, K)$.

Note that while our classifier is trained on alias sets identified by alias resolution techniques with known limitations, it is nonetheless able to identify new alias sets. 
We argue that this is because the training set is sufficiently rich due to its size and random selection of interfaces, providing considerable diversity and heterogeneity of loss traces across aliases. Our evaluation in Sec.~\ref{sec:evaluation} confirms this observation and confirms the ability of our technique to generalize patterns in the training dataset, i.e., the fourth aforementioned challenge.
\begin{table}[t]
\RawFloats
\centering
\parbox{.58\linewidth}{
\resizebox{\linewidth}{!}{%
\begin{tabular}{|l|r|r|r|r|r|r|}
\cline{2-7}
   \multicolumn{1}{l|}{} & \multicolumn{3}{|c}{IPv4} & \multicolumn{3}{|c|}{IPv6} \\
\cline{2-7}
    \multicolumn{1}{l|}{} & \makecell{Precision} & \makecell{Recall } &
    \makecell{F1 score}& \makecell{Precision}& \makecell{Recall}& \makecell{F1 score}\\
\hline
     \makecell{Random \\forest }& 0.990 & 0.499 & 0.652& 0.992 & 0.647 & 0.782  \\
\hline
	\makecell{Multilayer \\perceptron }& 0.993 & 0.431& 0.591
& 0.978 & 0.641 &0.769  \\
\hline
	\makecell{KNN} &0.952 & 0.638 & 0.764 & 0.970 & 0.622 & 0.756 \\
\hline
	\makecell{SVM}  & 0.986 & 0.478 & 0.642 & 0.988  & 0.599 & 0.743 \\
\hline
\end{tabular}}
\caption{Classifier performance on our test set averaged over ten training/testings.
}
\label{table:choose-classifier}
}
\quad
\parbox{.37\linewidth}{
\resizebox{\linewidth}{!}{%
\begin{tabular}{|l|l|l|}
\cline{2-3}
   \multicolumn{1}{l|}{} &  \multicolumn{2}{l|}{Gini index} \\
\hline   
Feature  & IPv4 & IPv6 \\
\hline
\makecell[l]{loss rate for the candidate $c$} & 0.169 & 0.192 \\
\hline
\makecell[l]{\texttt{burst} $\rightarrow$ \texttt{burst} transition \\ probability for the candidate $c$}  & 0.113 & 0.125 \\
\hline
\makecell[l]{\texttt{burst} $\rightarrow$ \texttt{burst} transition \\ probability for the seed $s$} &  0.101  & 0.121 \\
\hline
\makecell[l]{Pearson correlation coefficient} & 0.091 & 0.109 \\
\hline
\makecell[l]{loss rate for $\kappa_c$, \\ the control of the candidate $c$} & 0.077  & 0.104 \\
\hline
\end{tabular}}
\caption{The five most important features of our random forest classifiers.
}
\label{table:feature-importance}
}
\end{table}

\subsubsection{Choice of classifier.}
\label{sec:machine-learning-classifier}
We compared the performance of four classifiers that \texttt{scikit-learn} library offers, namely random forest, multilayer perceptron, $k$-nearest neighbors (KNN), and support vector machines (SVM). 
To this end, we evenly divided our dataset into a training and a test set, and compared these classifiers based on their precision, recall, and F1 score for both IPv4 and IPv6 datasets. 
Since \texttt{true} labels are only provided from aliases identified by \textsc{Midar} and Speedtrap,  the recall values  correspond to the portion of pairs of aliases in our training set 
that are detectable by both \textsc{Midar} and \tool (IPv4) or by both Speedtrap
and \tool (IPv6).
Table~\ref{table:choose-classifier} presents the averaged result of this comparison after performing 10 randomized splits of the training and test sets. All classifiers exhibit relatively good performance. 
We have decided to use the random forest classifier, which is composed of 500 trees, as it has the highest precision for both IPv4 and IPv6, and the best F1 score for the IPv6 dataset.

Finally, Table~\ref{table:feature-importance} shows the five most important features of our random forest classifiers based on the Gini index~\cite{cart84} that describes the weight of individual features in the classifier's decision. This table reveals a few important points. First, no single feature dominates the classifier's decision, particularly for IPv6. 
This confirms the complexity of the patterns for relating loss traces of aliases, as they cannot be accurately learned by a small number of features or simple threshold-based rules. Second, this table also illustrates that most of our engineered features are indeed very important in distinguishing loss traces of aliases. 
Third, the use of $\kappa_c$ as one of the main features suggests that the classifier distinguishes losses related to rate limiting from other losses.

\subsection{Refining the alias set}
\label{sec:robustness}
Independent network loss could accidentally result in classifying unrelated interfaces as aliases, i.e., generating false positives. To reduce the chance of this, \tool incorporates a refinement step,
\texttt{refine($A_s$)}, that involves repeating  \texttt{signature()} and
\texttt{classify()} on the previously-identified alias set $A_s$. If a candidate $c$ fails to be (re)classified as an alias of the seed $s$, it is removed from the alias set. This step is repeated 
until the alias set remains unchanged over two iterations. Sec.~\ref{sec:evaluation} evaluates the resulting reduction of false positives.

%

\section{Evaluation}
\label{sec:evaluation}
We evaluate \tool with regards to its ability (i) to identify alias pairs that state-of-the-art techniques, namely \textsc{Midar} and Speedtrap, are unable to identify, and (ii) to maintain a low rate of false positives. 

\subsubsection{Dataset.}
We evaluate \tool on  ground truth data from the Internet2 and \textsc{Switch} networks.
For Internet2, router configuration files were obtained
on 10 April, with measurements conducted on 11 and 12 April 2019.
There were 44 files, each corresponding to a single router. All are Juniper routers.
The files concern 985 IPv4 and 803 IPv6 addresses/interfaces, from
which we removed 436 IPv4 addresses and 435 IPv6 addresses that did
not respond to any probes sent by either \textsc{Midar}, Speedtrap, or
\tool. The resulting dataset consists of 6,577 IPv4 and 2,556 IPv6 alias pairs. For \textsc{Switch}, a single file was obtained on 3
May, with measurements conducted 3-5 May 2019.  The file
identified 173 Cisco routers running either IOS or IOS-XR.  From the
1,073 IPv4 and 706 IPv6 addresses listed in the file, we removed 121 
IPv4 and 29 IPv6 unresponsive addresses. 
The resulting dataset consists of 4,912 IPv4 and 2,641 IPv6 alias pairs.

\subsubsection{Reducing false positives.}
We computed the distribution of number of rounds for $\mathtt{refine}()$ to finalize the alias set for each seed in our dataset:
For 79\% (98\%) of all seeds, $\mathtt{refine}()$ takes 2 (3) more rounds. Note that the minimum of two rounds is required by design (Sec.~\ref{sec:robustness}) This basically implies that $\mathtt{refine}()$ only changed the alias set for 20\% of the seeds in a single round.

\subsubsection{Results.}
\setlength{\intextsep}{-5ex} 
\begin{figure}[t]
\RawFloats
\begin{floatrow}
\parbox{.6\linewidth}{
\capbtabbox{%

\resizebox{\linewidth}{!}{%
\begin{tabular}{|l|l|l|l|l|l|l|l|}
\cline{3-8}
    \multicolumn{2}{l|}{} &  \multicolumn{3}{c|}{IPv4} &   \multicolumn{3}{c|}{IPv6} \\
\cline{3-8}
    \multicolumn{2}{l|}{} & \textsc{Midar} & \makecell{ltd ltd} & \makecell{\textsc{Midar} \\$\cup$ \\ltd ltd } & Speedtrap & \makecell{ltd ltd} & \makecell{Speedtrap \\ $\cup$\\ ltd ltd}\\
\hline
    \multirow{2}{*}{Internet2} & \makecell[l]{Precision} & 1.000 & 1.000  & 1.000 & N/A  &1.000 & 1.000\\
\cline{2-8}
	 & \makecell[l]{Recall} & 0.673  & 0.800  & 0.868 & N/A & 0.684 & 0.684\\
\hline
    \multirow{2}{*}{\textsc{Switch}} & \makecell[l]{Precision} & 1.000 & 1.000  & 1.000 & 1.000  &1.000 & 1.000 \\
\cline{2-8}
	 & \makecell[l]{Recall} & 0.090  &  0.499  & 0.599  & 0.384 & 0.385 & 0.772\\
\hline
         
\end{tabular}}}
{%
  \caption{Evaluation on ground truth networks.}%
  \label{table:internet2-results}
}
}

\end{floatrow}
\end{figure}

Table~\ref{table:internet2-results} presents the precision and recall of \textsc{Midar}, Speedtrap, \tool, and the union of both tools on IPv4 and IPv6 ground truth data from the Internet2 and \textsc{Switch} networks. Note that it is possible for recall from the union of both tools to be greater than the sum of recall values for individual tools, as we observe in the \textsc{Switch} results.
This arises from the transitive closure of alias sets identified from the two tools that leads to the detection of additional alias pairs. The main findings of Table~\ref{table:internet2-results} can be summarized as follows:
\begin{enumerate}
\item \tool exhibits a high precision in identifying both IPv4 and
  IPv6 alias pairs from both networks with zero false positives.
\item \tool can effectively discover IPv6 aliases that state-of-the-art Speedtrap is unable to find. 
In the Internet2 network that uses Juniper routers,
\tool was able to identify 68.4\% of the IPv6 alias pairs while Speedtrap was unable to identify any.
In the \textsc{Switch} network that deploys Cisco routers, 
\tool\ and Speedtrap show comparable performance by identifying 38.5\% and 38.4\% of the IPv6 alias pairs,
respectively. The results were complementary, with the two tools together identifying 77.2\% of the IPv6 alias pairs, a small boost beyond simple addition of the two results
coming from the transitive closure of the alias sets found by each tool.

\item \tool\ can discover IPv4 aliases that state-of-the-art \textsc{Midar} is unable to find.
In the Internet2 network, \tool\ identifies 80.0\% while \textsc{Midar} detects 67.3\% of aliases.
In the \textsc{Switch} networks, \tool identified 49.9\% while \textsc{Midar} detects only 9.0\% of all aliases.
\end{enumerate}


A couple of detailed observations follow.
We conducted follow up analysis on the behavior of Speedtrap and \textsc{Midar} to ensure proper assessment of these tools. 
First, we examined Speedtrap's logs to diagnose Speedtrap's inability to detect any IPv6 aliases for Internet2. We noticed that every fragmentation identifier time series that Speedtrap seeks to use as a signature, was either labeled as random or unresponsive. This was not surprising, as prior work on Speedtrap~\cite{luckie2013speedtrap} also reported that this technique does
not apply to the Juniper routers that primarily comprise Internet2.
Second, we explored \textsc{Midar}'s logs to investigate the cause of its low recall for \textsc{Switch}. We learned that only one third of the IPv4 addresses in this network have monotonically increasing IP IDs.

\subsubsection{Limitations and future work.}
Because \icmprl could be triggered at thousands of packets per second, \tool requires the sending of many more packets than other state-of-the-art alias resolution techniques. The maximum observed probing rate during the experiments for this paper was 34,000 pps from a single vantage point during a 5-second round. On Internet2 (\textsc{Switch}), \textsc{Midar} and Speedtrap sent 164.5k (106k) and 4k (12.7k) probe packets while \tool sent about 4,8M (12.7M) packets.
In future work, we plan to explore ways to reduce the overhead
of probing and make \tool more scalable.

\section{Ethical Considerations}\label{sec:ethical}
\tool works by triggering limits in routers that are there for protective reasons. This raises ethical concerns, which we discuss below. To evaluate the impact of \tool, we have taken two steps: experiments in a lab environment (Sec.~\ref{sec:lab-tests} and Appendix~\ref{sec:ethical-appendix}), and feedback from operators (Sec.~\ref{sec:operator-feedback}).
\subsection{Lab experiments}\label{sec:lab-tests}
We have run experiments in a lab environment on conservatively chosen hardware (over 10 years old) to show that  
\tool has a controlled impact. Our findings are that:
(1) routers being probed with Echo Requests by the tool remain reachable to others via
\texttt{ping} with a high probability; 
and (2) Router CPUs show a manageable overhead at the highest probing rate, leading us to believe that our measurements are unlikely to impact the control and data planes.
(3) Both \tool and existing measurement techniques impact troubleshooting efforts (e.g., \texttt{ping}, \texttt{traceroute}). \tool does not
stand out in terms of impact compared with other accepted techniques.
Appendix~\ref{sec:ethical-appendix} details the experiments
which support these conclusions.
\subsection{Real-world operator feedback}\label{sec:operator-feedback}
In addition to lab experiments, we conducted joint experiments
with
\textsc{SURFnet} and \textsc{Switch} to evaluate the potential impact of \tool.
The experiment consisted in running \tool on their routers while they were
monitoring the CPU usage.
Each run lasted about 1 minute.
For \textsc{SURFnet}, we ran \tool on two Juniper routers: an MX240 and an MX204. 
The operator observed a 4\% and 2\% CPU overhead. 
The operator also told us that the CPU overhead was observed on the MPC (line modules) CPU and not the central routing engine CPU.
For \textsc{Switch}, we ran \tool on three Cisco routers: 
an NCS 55A1, an ASR 9001, and an ASR-920-24SZ-M.
On the two first routers, the operator told us that there was no observable change in CPU utilization.
On the third router, which has a lower CPU capacity than the two others, the operator observed a CPU overhead up to 29\%.
These results confirm our belief that \tool is unlikely to impact the control and data planes.

\section{Conclusion}\label{sec:conclusion}
This paper presents \tool, a new, high-precision alias resolution
technique for both IPv4 and IPv6 networks that leverages the \icmprl feature of individual routers.  
We have shown that \icmprl can generate loss traces that can be used to reliably identify aliases from other interfaces. 
\tool enables IPv6 alias resolution on networks composed of Juniper routers that the state-of-the-art Speedtrap technique is not able to identify.  
As a part of our future work, we plan to enhance the efficiency of \tool and explore the use of \icmprl for fingerprinting individual routers.
Both the source code for \tool and our dataset are publicly available\footnote{https://gitlab.planet-lab.eu/cartography}.

\section*{Acknowledgments}
We thank Niels den Otter from \textsc{SURFnet} and Simon Leinen from \textsc{Switch} network for their time in conducting joint experiments of \tool.
We thank people from Internet2 and \textsc{Switch} for providing 
the ground truth of their network.
We thank
the anonymous
reviewers from both the PAM TPC and our shepherd, for their
careful reading of this paper and suggestions for its improvement.
Kevin Vermeulen, Olivier Fourmaux, and Timur Friedman are associated with
Sorbonne Université, CNRS, Laboratoire d'informatique de Paris 6, LIP6,
F-75005 Paris, France. Kevin Vermeulen and Timur Friedman are associated
with the Laboratory of Information, Networking and Communication Sciences, LINCS,
F-75013 Paris, France.
A research grant from the French Ministry of Defense has made this
work possible.

\appendix
\section{Ethical considerations}\label{sec:ethical-appendix}

\subsection{Precautions taken.}
We take two precautions, that we understand to be community best practice:
We sent all probing traffic from IP addresses that were clearly
associated via \textsc{WhoIs} with their host locations, either at our
institution or others hosting PlanetLab Europe nodes. 
We have also set up a web server on the probing machines
with a contact email, so that any network operators could opt out
from our experiment.
We received no notice whatsoever from network operators expressing
concern about our measurements. Though this is a positive sign, it
could be that there are impacts that were not noticed, or that the
concerns did not reach us. We therefore pushed our examination
further, as detailed in the following sections.

\subsection{Impact on other measurements.}
\begin{figure}[t]
\centering     
\subfigure[For $\mathtt{find\_rate}()$]{\includegraphics[width=.4\textwidth]{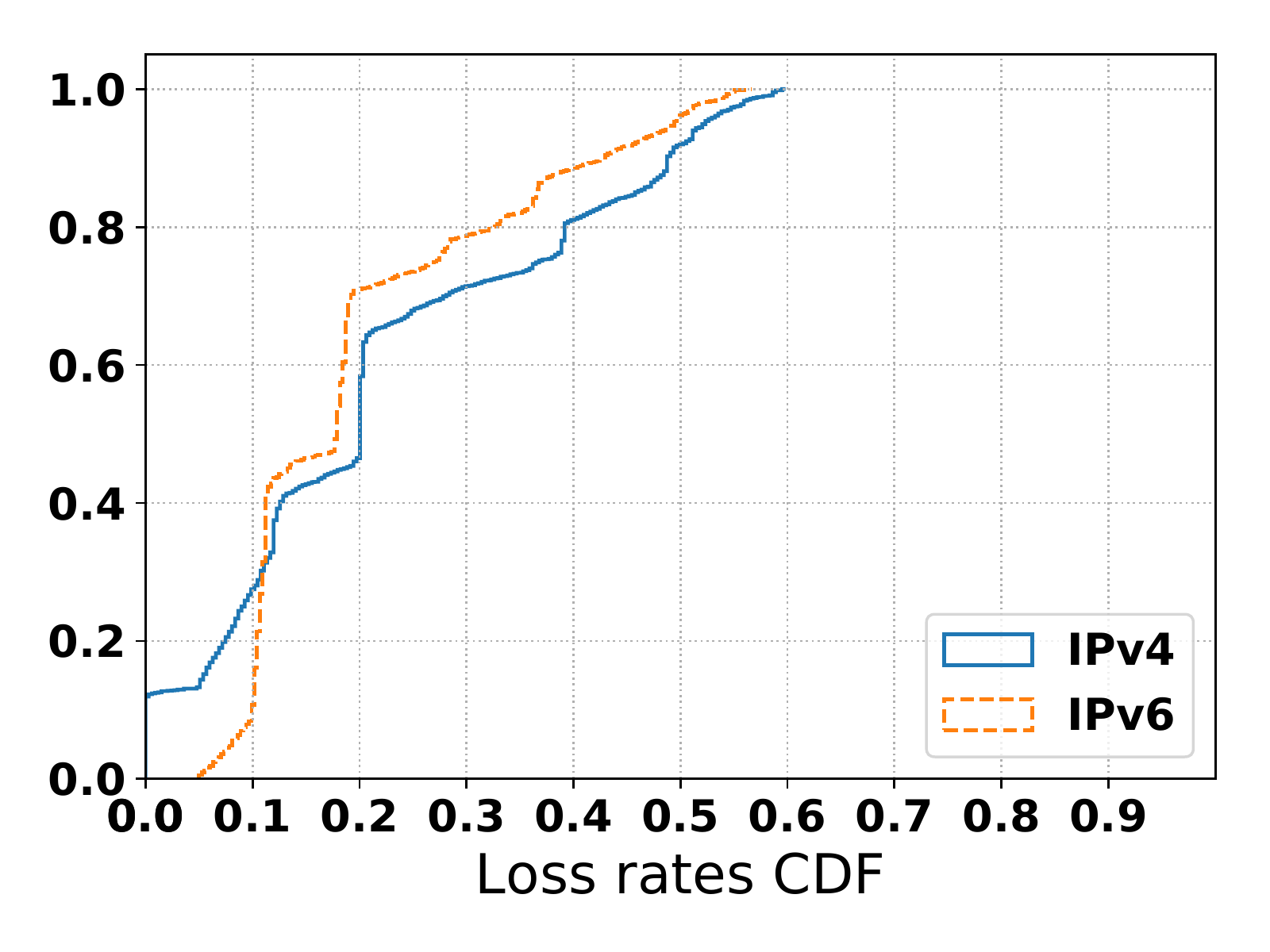}}%
\subfigure[For $\mathtt{signatures}()$]{\includegraphics[width=.4\textwidth]{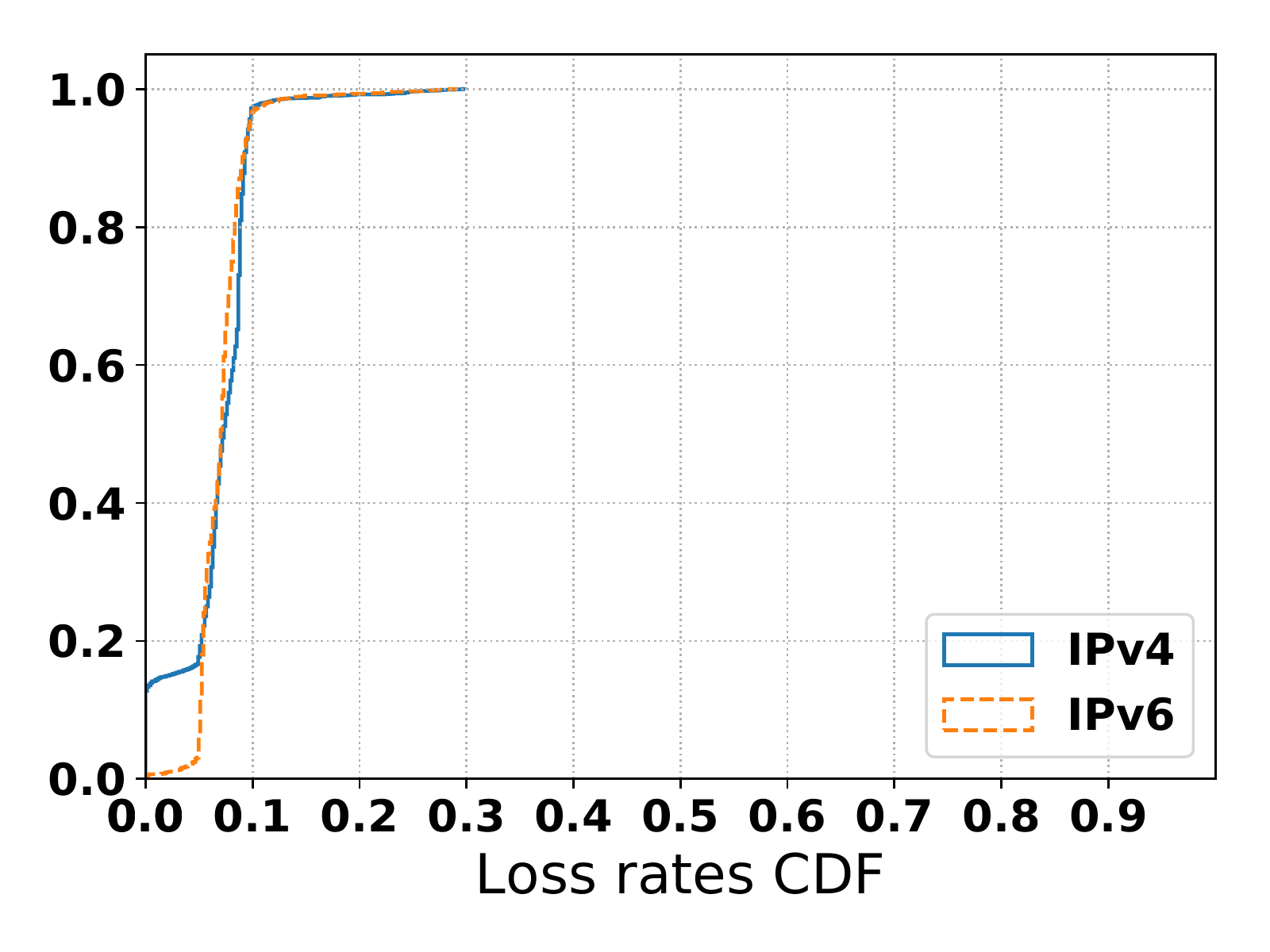}}
\caption{Maximum loss rates}
\label{fig:loss-rates}
\end{figure}
\tool's $\mathtt{find\_rate}()$ aims to find an ICMP
Echo Request probing rate that produces an Echo Reply trace with a
loss rate in the $[0.05,0.10]$ range. While it is searching for this rate,
it can induce a loss rate above 0.10. If it does so, it proceeds to
a binary search to find a lower probing rate for which traces
falls within the desired range. Fig.~\ref{fig:loss-rates} shows
that loss rates can go as high as 0.60.

The impact on reachability for the IP addresses of that node is that
there is a worst case 0.60 probability that a single \texttt{ping} packet to
such an address will not receive a response if it arrives at the node
during the five seconds of highest rate probing time. Most pings occur
in series of packets, so the worst case probabilities are 0.36 for two
\texttt{ping} packets being lost, 0.22 for three, 0.13 for four, 0.08 for five,
and 0.05 for six. These are worst case probabilities for the five
seconds at highest loss rate. Average reachability failure
probabilities are 0.22 for one \texttt{ping} packet, 0.05 for two, 0.01 for
three, and so on, while a node is being probed at its highest rate.
To judge whether such a level of interference with other measurements
is exceptional, we compare it to the impact of the state-of-the-art
\textsc{Midar} tool. \textsc{Midar} has a phase during which it
elicits three series of 30 responses each, using different methods for each
series: TCP SYN packets, to elicit TCP RST or TCP SYN-ACK responses;
UDP packets to a high port number, to elicit ICMP Destination
Unreachable responses; and ICMP Echo Request packets, to elicit ICMP
Echo Reply responses~\cite{keys2013internet}. The probing rate is very
low compared to \tool: a mere 100 packets per second across
multiple addresses. This is not a concern for the TCP and ICMP
probing. However, the UDP probing taps into an \icmprl
mechanism that tends to be much less robust than the typical ICMP Echo
Reply mechanism on some routers. ICMP Destination Unreachable messages are often rate limited
at 2 packets per second, which is $1/500^\mathrm{th}$ the typical
rate at which ICMP Echo Reply messages are rate limited. (For example,
the default rate at which Cisco routers limit ICMP Destination
Unreachable messages is 1 every 500 ms.)
\begin{figure}[t]
\centering     
\subfigure[Before and after \textsc{Midar} run]{\includegraphics[width=.5\textwidth]{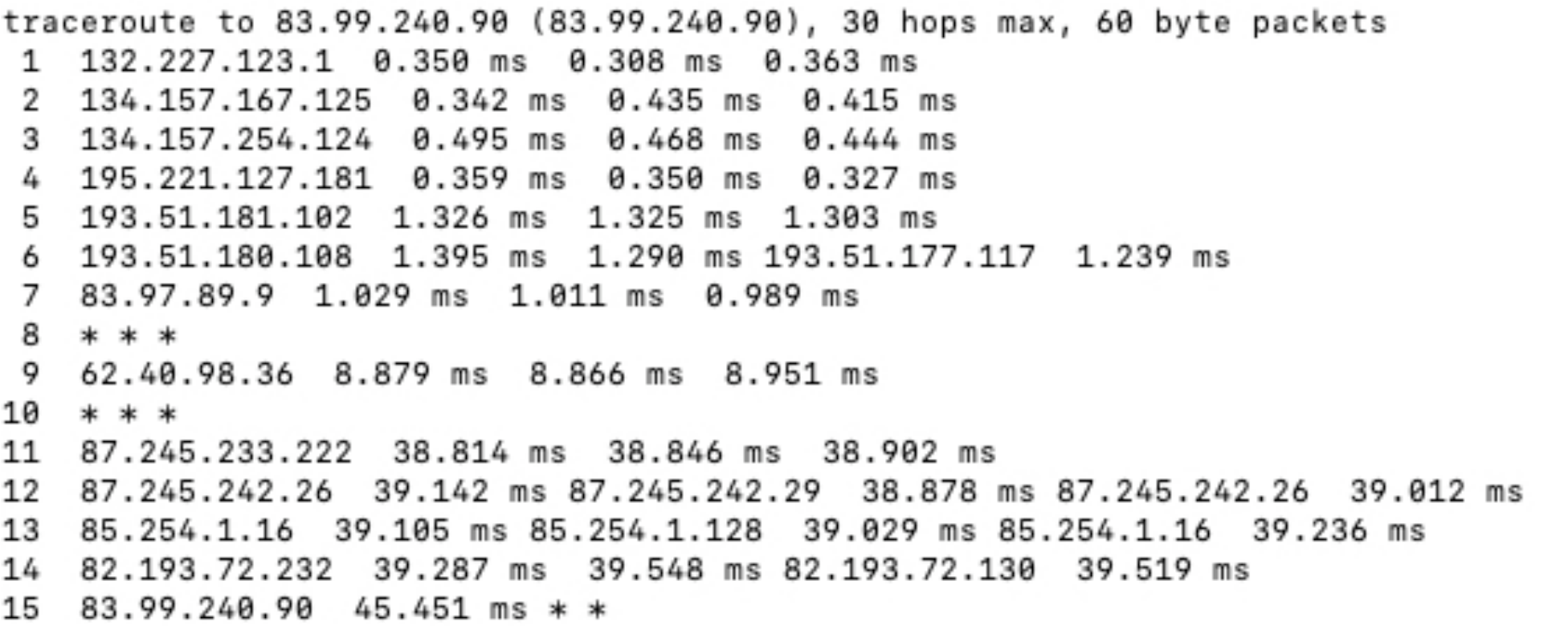}}%
\subfigure[While \textsc{Midar} is running]{\includegraphics[width=.5\textwidth]{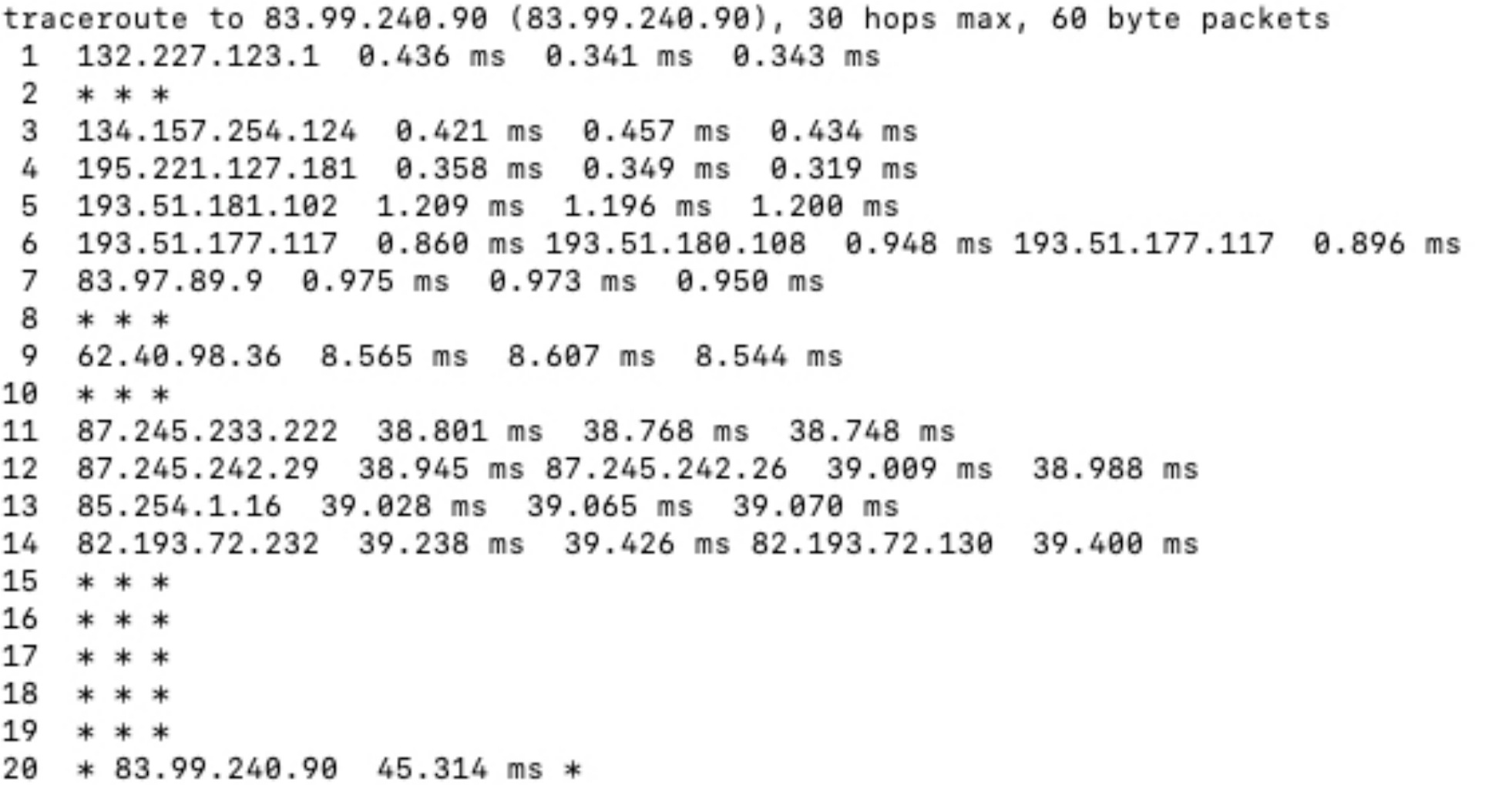}}
\caption{Example erroneous \texttt{traceroute} result}
\label{fig:midar-impact}
\end{figure}

We found that, when an IP address is a \texttt{traceroute}
destination, \textsc{Midar} can completely block ICMP Destination
Unreachable messages coming from that
destination. 
Fig.~\ref{fig:midar-impact} illustrates the impact. 
The
figure shows two \texttt{traceroute} results, the top one from before
or after \textsc{Midar} being run, and the bottom one during
\textsc{Midar} probing. During the \textsc{Midar} run, we see that
\texttt{traceroute} receives no responses while it is probing hop 15,
where the destination is in fact to be found. The normal functioning
of \texttt{traceroute} is to continue probing at higher and higher hop
counts. Only a few seconds later, when \texttt{traceroute} is sending
probes to hop 20, does it start to receive ICMP Destination
Unreachable messages from the destination. 
The result is an erroneous
\texttt{traceroute}, indicating that the destination is five hops
further away than it actually is. We observed this erroneous
\texttt{traceroute} effect on 2,196 IP addresses out of a dataset of 10,000
IPv4 addresses collected from across the Internet.
For
both \tool\ and \textsc{Midar}, 
transient interference with
other measurements can be observed for the few seconds during which an
IP address is being probed.
Our conclusion is not that the diminution in \texttt{ping}
reachability induced by \tool\ is necessarily anodyne. Care
should be taken to circumscribe this effect. 
But we observe that it
does not stand out in terms of its impact on other measurements.

\subsubsection{CPU usage.}
We now examine the CPU overhead generated by \tool, and its
potential impact on the forwarding plane and other features involving
the CPU.  
We have run an experiment in a local network with our own
Cisco (model 3825, IOS 12.3) and Juniper (model J4350, JunOS 8.0R2.8) routers.
The experiment consists in measuring three metrics while $\mathtt{find\_rate}()$ routine of \tool, which has the highest probing
rate, is running.
We measured: (1) The CPU usage of the router, (2)
the throughput of a TCP connection between the two end hosts, and (3) the rate of BGP updates.
\icmprl is configured on both our Juniper and Cisco
routers with an access list~\cite{juniper-rate-limiting-acl,cisco-acl}, 
limiting the ICMP input bandwidth destined to the
router to 1,000 packets per second, which is the default
configuration on Juniper routers.

TCP throughput was unaffected, at an average of 537 Mbps and BGP
updates remained constant at 10 per second.
CPU usage was at 5\% for Cisco and 15\% for Juniper when \tool\ was not probing.
During the probing, 
the maximum
overhead was triggered for both at a maximum probing rate of 2,048
packets per second, with a peak at 10\% for Cisco and 40\% for Juniper
during 5 seconds. Our conclusion is that there is an impact of
high probing rates on CPU, but we do not witness a disruptive impact
on either the data plane (TCP throughput) or the control plane (BGP
update rate).

 \bibliographystyle{splncs04}
 \bibliography{bibliography-main-ccr}

\end{document}